\documentclass[manuscript,nonacm]{acmart}

\usepackage{subcaption}

\AtBeginDocument{%
  }

\setcopyright{cc}
\setcctype{by-sa}
\copyrightyear{2025}

\begin{document}

\title{Inductive Loop Analysis for Practical HPC Application Optimization}

\author{Philipp Schaad}
\orcid{0000-0002-8429-7803}
\email{philipp.schaad@inf.ethz.ch}
\affiliation{%
  \institution{ETH Zurich}
  \city{Zurich}
  \country{Switzerland}
}

\author{Tal Ben-Nun}
\orcid{0000-0002-3657-6568}
\email{talbn@llnl.gov}
\affiliation{
    \institution{Lawrence Livermore National Laboratory (LLNL)}
    \city{Livermore}
    \country{USA}
}

\author{Patrick Iff}
\orcid{0000-0001-5979-4915}
\email{patrick.iff@inf.ethz.ch}
\affiliation{
    \institution{ETH Zurich}
    \city{Zurich}
    \country{Switzerland}
}

\author{Torsten Hoefler}
\orcid{0000-0002-1333-9797}
\email{htor@inf.ethz.ch}
\affiliation{
    \institution{ETH Zurich}
    \city{Zurich}
    \country{Switzerland}
}

\renewcommand{\shortauthors}{Schaad et al.}

\begin{abstract}
Scientific computing applications heavily rely on multi-level loop nests operating on multidimensional arrays.
This presents multiple optimization opportunities from exploiting parallelism to reducing data movement through prefetching and improved register usage.
HPC frameworks often delegate fine-grained data movement optimization to compilers, but their low-level representations hamper analysis of common patterns, such as strided data accesses and loop-carried dependencies.
In this paper, we introduce symbolic, inductive loop optimization (SILO), a novel technique that models data accesses and dependencies as functions of loop nest strides.
This abstraction enables the automatic parallelization of sequentially-dependent loops, as well as data movement optimizations including software prefetching and pointer incrementation to reduce register spills.
We demonstrate SILO on fundamental kernels from scientific applications with a focus on atmospheric models and numerical solvers, achieving up to 12$\times$ speedup over the state of the art.
\end{abstract}



\begin{teaserfigure}
    \centering
    \includegraphics[width=\linewidth]{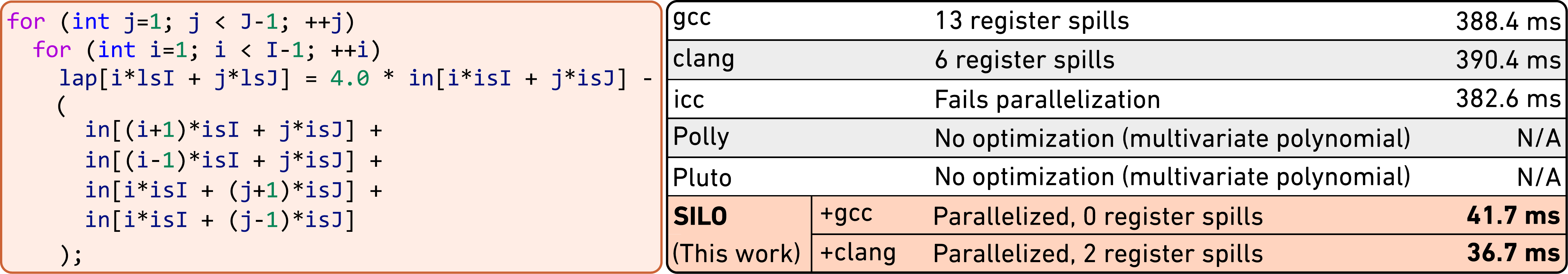}
    \caption{Data accesses with parametric strides in a 2D Laplace operator cause missed optimization opportunities with modern optimizing compilers on an 18-core Intel Xeon Gold 6140.}
    \Description{A code sample shows a 2D Laplace operator implemented in C. To the right, a table shows how existing optimizing compilers and HPC frameworks fail to adequately optimize the provided code due to parametric strided data accesses. SILO (this work) successfully optimizes the code by parallelizing and reducing the number of register spills.}
    \label{fig:posterchild}
\end{teaserfigure}
\maketitle

\section{Introduction}
Scientific computing applications, deep neural networks, and numerical weather prediction systems heavily rely on loops, with the majority of their computation time spent within them.
Consequently, loops are central to optimization strategies in High Performance Computing (HPC) applications.
In the post-Moore era a lot of focus is put on maximizing parallelism and improving data locality, for which loops are no exception.
Various techniques are employed to achieve this, including vectorizing, pipelining, or tiling of loops, or changing of data layouts.

These optimizations are typically carried out by loop nest optimizers, which analyze loops holistically as groups of operations to be rescheduled.
For instance, state-of-the-art optimizers frequently utilize the polyhedral model~\cite{pop2006graphite, merouani_looper_2024, bondhugula_pluto_2016, grosser_polly-polyhedral_2011, grosser_pollyperforming_2012, zinenko_visual_2018, chen_chill_2008} which relies on counting points in polytopes that represent loop nests.
While effective for optimizing loop schedules, this approach depends on well-formed loops and affine index memory accesses.
These are assumptions that frequently do not hold in practical HPC applications involving complex data structures and custom-padded multidimensional arrays.
As illustrated in Fig.~\ref{fig:posterchild}, common patterns such as linearized data accesses lead to missed optimization opportunities in both general-purpose compilers and polyhedral optimizers.

Domain Specific Language (DSL) toolchains~\cite{ragan-kelley_halide_2013, paredes_gt4py_2023, gysi_stella_2015, tang_pochoir_2011, baghdadi_tiramisu_2018, chen_tvm_2018} address this challenge by providing loop abstractions that facilitate effective analysis and rescheduling.
However, these approaches are inherently limited to specific application domains.
Moreover, while loop schedule adaptations significantly improve data locality, further memory optimizations, such as array privatization, can expose substantial additional parallelism.

In this work, we propose \emph{inductive loop analysis}: a symbolic dependency analysis across successive iterations that enables a novel approach to loop optimization for real-world HPC applications.
Our technique, \textbf{S}ymbolic, \textbf{I}nductive \textbf{L}oop \textbf{O}ptimization (SILO), models data accesses within loops as injective functions over loop variables, focusing exclusively on the stride between successive loop iterations.
This abstraction supports inductive reasoning about loop behavior and dependencies, even in the presence of varying strides and complex, non-affine data access patterns.

Building on inductive loop analysis, we present practical optimization techniques that complement the optimizations performed by general-purpose compilers and existing optimization frameworks.
By simplifying dependency analysis across successive iterations, SILO enables parallelism extraction through array privatization and automatic pipeline parallelization, thereby maximizing hardware utilization.
Beyond loop schedule optimization, we also target \emph{memory schedules}, reducing data movement costs in loops with complex access patterns.
Symbolic access order information is used for generating application-informed software prefetching instructions and to minimize register spills inside loops.

We implement a prototype of SILO on an existing compiler infrastructure~\cite{ben-nun_stateful_2019} and use a series of fundamental kernels from scientific computing to showcase the real-world applicability of our techniques.
Our results show that SILO can be used in coordination with the optimizations performed by HPC optimization frameworks and compilers to achieve consistent performance improvements over the state of the art.

In summary, we make the following contributions:
\begin{itemize}
    \item An inductive loop analysis technique through symbolic dependence analysis of successive iterations
    \item Inductive analysis guided optimizations, improving parallelism, register allocation, and memory prefetching
    \item A proof-of-concept implementation yielding up to 12$\times$ speedup in a real-world applications and HPC benchmarks
\end{itemize}

\section{Symbolic Inductive Loop Optimization}
Scientific computing applications rely not just on loops, but multi-level loop nests.
Loop nests are a collection of arbitrarily many individual loops nested inside each other, each containing an arbitrary number of program statements operating on some data.
Often these loops are not perfectly nested, meaning there are potentially many statements between individual loop levels.
Data accesses inside a loop may be reads and writes to scalar values, but more commonly for HPC applications, loops operate on arrays.
These arrays can have multiple dimensions, and accesses are characterized by some pointer to the data container, together with an offset expression.
These offset expressions often depend on the loop variables of one or more loops in a nest or even on other data.
Additionally, individual loop bounds or step sizes may themselves depend on the loop variables of other loops around them, or even their own loop variable, such as the loop nests shown in Fig.~\ref{fig:changing-stride}.

\subsection{Characterizing Loops and Data Accesses Inductively}
To reason about data accesses and dependencies in this complex environment without performing significant over-approximations, we first establish a framework around loops and their data accesses built on symbolic expressions.
We characterize any analyzable loop $\mathcal{L}$ by four parameters and the loop body $\mathcal{L}_{body}$.
The four parameters are the loop variable or symbol $\mathcal{L}_{var}$, and three symbolic expressions $\mathcal{L}_{start}$, $\mathcal{L}_{end}$, and $\mathcal{L}_{stride}$.
These three expressions dictate the starting value of the loop symbol $\mathcal{L}_{var}$, the symbol's value after the last loop iteration, and the loop stride, respectively.
Each symbolic expressions can be treated as mathematical functions, injective w.r.t. the \textit{current} loop variable, which operate on program parameters that do not change over the course of the loop, loop iteration variables, or other symbols in the loop nest.

The loop body $\mathcal{L}_{body}$ is characterized as a collection of one or more arbitrary program statements.
Each program statement is represented by a set of data reads or consumed elements, and a set of writes or produced elements.
These reads and writes establish a set of data dependency relationships between program statements.
Each read and write is represented by the name of (or a pointer to) a data container $D$ and a symbolic expression $f$ determining the offset to the start of the data container, denoted $D[f]$.
Similar to the symbolic expressions characterizing a loop, $f$ is an injective function on symbolic parameters, such as the loop variable $\mathcal{L}_{var}$ or other program symbols and parameters.

\begin{figure}
    \centering
    \includegraphics[width=.6\linewidth]{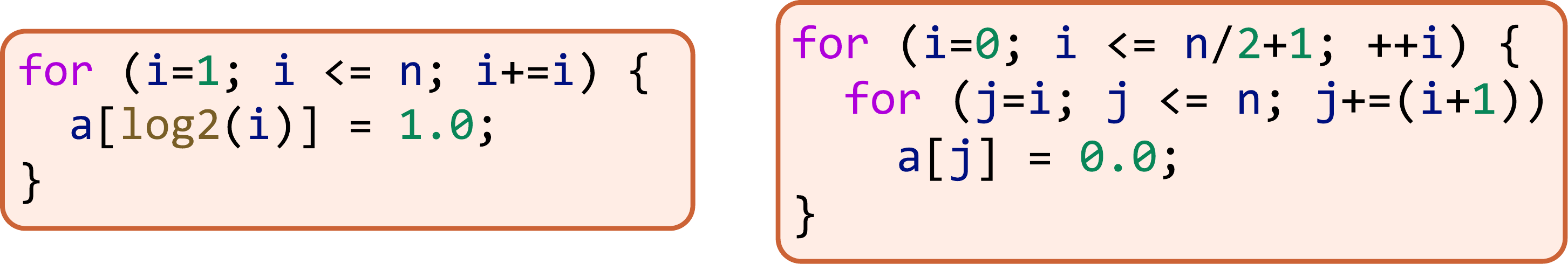}
    \vspace{-.3cm}
    \caption{Variable loop strides cause missed optimization opportunities, even in modern polyhedral compilers.}
    \Description{Two code samples show how variable loop strides can lead to missed optimization opportunities. In the first code sample, the loop stride is given by a variable i, which is doubled in each loop iteration. In the second code sample, there are two nested loops, where the inner loop's iteration variable is incremented by the outer loop's iteration variable.}
    \vspace{-.3cm}
    \label{fig:changing-stride}
\end{figure}

For example, the loop shown on the left side of Fig.~\ref{fig:changing-stride} with an iteration variable $\mathcal{L}_{var} =$ \texttt{i} can be characterized with an expression $\mathcal{L}_{start}$ of a constant $0$, an expression $\mathcal{L}_{end}$ that represents some parameter $\leq n$, and an expression $\mathcal{L}_{stride}$ that is a function of $\mathcal{L}_{var}$ itself.
The write for the program statement in the loop body is expressed as a pointer to the array $a$ and a corresponding offset expression $f(i) = \log_2(i)$.
Similarly, the inner loop in the loop nest on the right side of Fig.~\ref{fig:changing-stride} has a stride expression $\mathcal{L}_{stride}$ that is a function of the outer loop's iteration variable \texttt{i}.
Because of these stride expressions that may take on changing values during iteration, neither loop nest is considered by polyhedral tools such as Polly~\cite{grosser_polly-polyhedral_2011} or Pluto~\cite{bondhugula_pluto_2016}.

With these basic elements we can summarize each loop and represent it as a single, black-box statement with a set of read and write data containers.
We discuss how and when these can be identified accurately without over-approximation in Section~\ref{ss:canonicalization}.
This allows us to represent entire loop nests as a hierarchy or tree of other loops and statements and reason about them inductively with symbolic arithmetic, without having to construct and analyze full iteration spaces.

\subsection{SILO Architecture}\label{ss:architecture}
\begin{figure}
    \centering
    \def\svgwidth{\linewidth}
    \includegraphics[width=.9\linewidth]{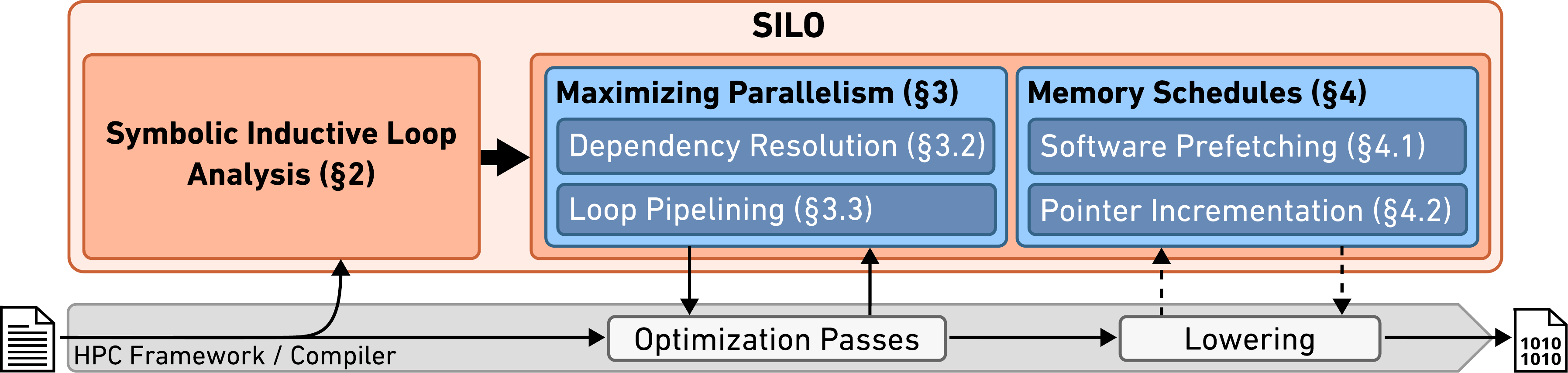}
    \vspace{-.3cm}
    \caption{An overview of the SILO architecture. Parallelism is improved through optimization passes in tandem with HPC frameworks' optimizations, while more fine-grained memory schedule optimizations are performed during lowering of the IR.}
    \Description{An overview of the architecture of SILO. SILO hooks into the optimization and IR lowering steps of an HPC optimizing compiler / framework to perform symbolic loop and data access analysis, before applying optimizations and lowering accordingly.}
    \label{fig:overview}
\end{figure}

SILO makes use of this symbolic representation of loops and data accesses to enrich the optimization capabilities of source-to-source program optimization frameworks and optimizing compilers.
It operates on the intermediate program representation (IR) used by an optimizing framework, which is expressive and high-level enough to retrieve the aforementioned symbolic expressions from loops and data accesses.
The symbolic loop and data access representation is used to optimize the IR by exposing more parallelism from seemingly sequential loops and by scheduling individual data accesses to improve prefetching and register usage.
We discuss both of these optimizations and the required analyses in detail in Sections~\ref{s:max-parallel} and \ref{s:memory-schedules}, respectively.
Optimizations are carried out in tandem with the HPC framework's optimization passes to let it make use of the removed sequential dependencies.
At the end of optimization, during lowering of the code to a different compiler representation or assembly, custom lowering rules help the HPC framework implement the selected memory schedules.
An overview of this architecture can be seen in Fig.~\ref{fig:overview}, and we discuss a practical implementation of it and the requirements for one in Section~\ref{s:implementation}.

\section{Maximizing Parallelism}\label{s:max-parallel}
With the majority of scientific computing application execution time spent inside loop nests, it is crucial that loop-level parallelism is exploited wherever possible.

Optimizing compilers such as Intel's icc/icx and polyhedral frameworks manage to extract a large degree of parallelism from loop nests in the absence of loop-carried dependencies.
Even certain sequential data dependencies can be parallelized with special tiling techniques for read-after-write dependencies in stencil computations~\cite{bondhugula_diamond_2017}.
However, these techniques only modify the loop schedule.
By also considering data allocation changes we can resolve data dependencies that may prevent more straight-forward and efficient parallelization strategies from being applicable.

The most common data dependencies are read-after-write (RAW) dependencies between loop iterations, where an iteration $i$ of a loop consumes a value produced by a previous iteration $i-x$ for some $x > 0$.
An inverse of that is a write-after-read (WAR) dependency, or input dependency, where a loop iteration $i$ reads from the same data address that a subsequent loop iteration $i+x$ will write to.
A third common dependency is a write-after-write (WAW) dependency, or output dependency, where subsequent loop iterations write to the same memory location.

\begin{figure}
    \centering
    \includegraphics[width=.95\linewidth]{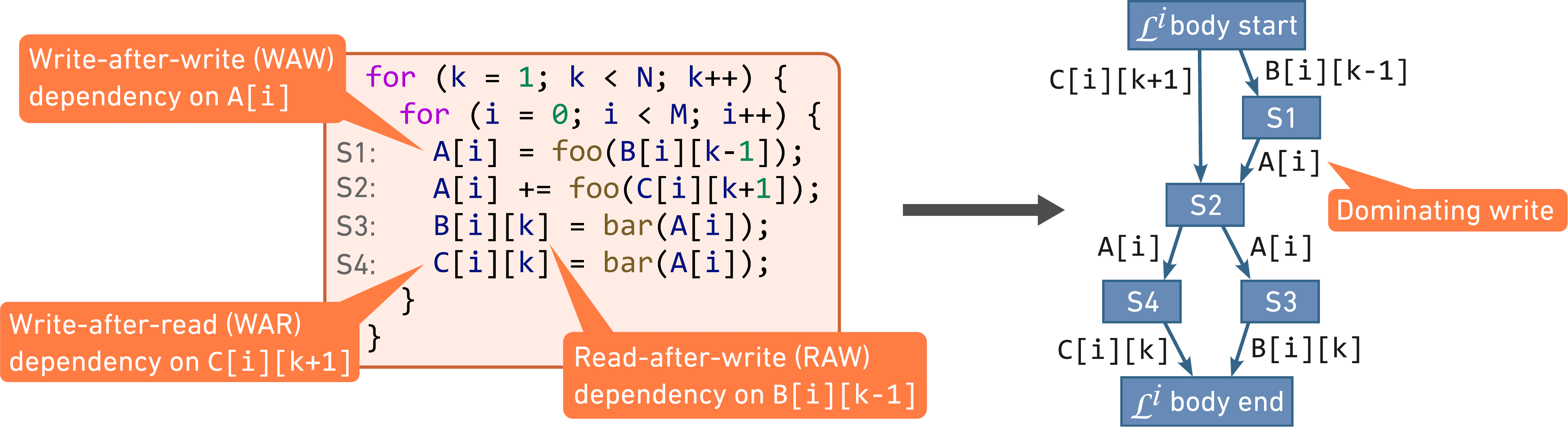}
    \caption{Loop-carried dependencies preventing parallelization of the \texttt{k}-loop. The dataflow graph of the \texttt{i}-loop body shows that reads from \texttt{A} are dominated by a write within the same iteration, potentially allowing for privatization.}
    \Description{A code sample shows a write-after-write conflict, a read-after-write conflict, and a write-after-read conflict, which prevent a loop nest from being parallelized. The image further shows a dataflow graph representation of the piece of code, where a write dominates one of the conflicting reads, indicating that parallelization may indeed be possible.}
    \label{fig:parallel-deps}
\end{figure}

We illustrate here how we can use inductive loop analysis with SILO to address some of these parallelization preventing dependencies using a didactic example shown in Fig.~\ref{fig:parallel-deps}.
The code shown on the left of Fig.~\ref{fig:parallel-deps} demonstrates a loop nest that may not be fully parallelized due to such data dependencies.
The inner loop $\mathcal{L}^i$ is fully data parallel, so one parallelization strategy is to move the outer loop $\mathcal{L}^k$ to the inside and then parallelize over $\mathcal{L}^i$.
However, if \texttt{M} is small w.r.t. \texttt{N}, this may underutilize the available hardware.

$\mathcal{L}^k$ on the other hand exhibits all three of the described data dependencies, preventing parallelization:
\begin{enumerate}
    \item RAW dependency on \texttt{B}, where each iteration consumes values produced in the previous iteration 
    \item WAR dependency on \texttt{C}, where an iteration reads a value overwritten by a later iteration
    \item WAW dependency on \texttt{A}.
\end{enumerate}

Most automatic parallelization techniques reliably detect the possibility of such dependencies and consequently preserve program correctness by not parallelizing.
However, in many cases such dependencies may be resolved through data access changes to allow subsequent parallelization.
We discuss in the following how SILO employs the symbolic loop and data access representation to identify and resolve common loop dependencies.

\subsection{Consumer and Producer Analysis}~\label{ss:canonicalization}
To analyze data dependencies between loop iterations, we initially analyze which data reads and writes of an individual loop iteration are externally visible.
The externally visible products of a loop iteration are all writes performed by it, with the exception of writes to data containers that do not live beyond a single iteration.
Externally visible consumed elements on the other hand are data reads inside the loop body, where the read value is not guaranteed to come from within the same loop iteration, and may hence be influenced by a different iteration.

We analyze these externally visible reads and writes by constructing a dataflow graph for the statements found in the body of the loop in question.
In this graph, data dependencies between statements are expressed with the data container in question and an offset into it given by a symbolic expression, as shown in Fig.~\ref{fig:parallel-deps}.
By traversing this graph, all writes are gathered into a list of externally visible writes, except if the definition of the data container is within the same loop body.
Similarly, all reads are collected into a list of external reads, except if the read is guaranteed to be self-contained.
A read with a specific offset expression $f$ to a data container $D$ is self-contained if there is a write to $D$ with a symbolically equivalent injective offset expression $g = f$ which dominates the read in the dataflow graph.

With the externally visible reads and writes for a single loop iteration, we may also infer the externally visible reads and writes of the loop as a whole.
We go over each external read and write of a single loop iteration and \emph{propagate} the corresponding array offset expressions based on the symbolic expressions of the loop's start- and endpoint and its stride.
Instances of the loop's iteration variable inside the offset expressions are given a specific range of values based on the range determined by the loop's characterizing expressions.
This gives us the exact externally visible reads and writes for the loop as a whole, through all iterations.
Where the iteration space of the loop is not statically countable through the given symbolic expressions, the access offsets are propagated to include the entirety of the data container to conservatively over-approximate.

\subsection{Eliminating Dependencies}
With the externally visible reads and writes of each loop nest determined, we can begin eliminating data dependencies.

\subsubsection{Removing External Writes}
A key to resolving output dependencies is removing as many writes to data containers with lifetimes beyond the scope of a single loop iteration as possible.
The goal is to identify instances where writes to and subsequent reads from loop external data containers can be replaced with a write and subsequent reads from a register.

This replacement can be performed whenever an externally visible write to a data container $D$ at offset $f$ is not read anywhere outside the loop in question, including the loop's own externally visible reads.
Checking if a data container is used anywhere else is readily attainable information in some optimization frameworks or compiler infrastructures~\cite{ben-nun_stateful_2019}, but otherwise requires a more detailed dataflow analysis.
This can be achieved by constructing a dataflow graph of the application surrounding the currently analyzed loop nest.
In this dataflow graph, each other loop nest can be represented as a single element with only externally visible reads and writes, hiding self-contained reads inside it.

\begin{figure}
    \centering
    \includegraphics[width=\linewidth]{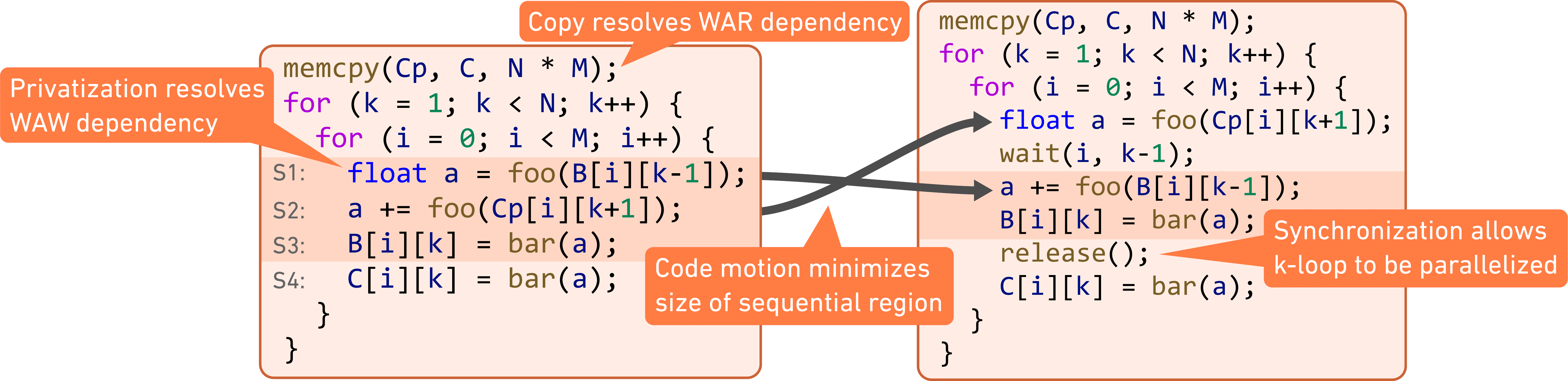}
    \vspace{-.5cm}
    \caption{Resolved write-after-write (WAW) and write-after-read (WAR) dependencies enable DOACROSS parallelization with synchronization if only read-after-write (RAW) dependencies prevent naive parallelization.}
    \Description{The code sample shows the previous code discussed, but with write-after-write and write-after-read dependencies resolved. With these dependencies resolved, the code sample shows how the code can now be parallelized using DOACROSS synchronization, and how code motion can be employed to minimize the sequential region by swapping statements S1 and S2.}
    \label{fig:doacross-parallelization}
    \vspace{-.3cm}
\end{figure}

The resulting directed dataflow graph is traversed and any read operation from the data container $D$ is a possible conflict that prevents privatization to a register.
To check whether a conflict is indeed present, the corresponding offset expression $g$ is checked against the propagated offset expressions for the externally visible write of the current loop.
If no intersecting offset expression can be found in the dataflow graph, the accesses inside the loop may be privatized to a register.
For the loop nest from Fig.~\ref{fig:parallel-deps} this corresponds to replacing accesses to \texttt{A[i]} with temporary values, as seen on the left-hand side of Fig.~\ref{fig:doacross-parallelization}.

\subsubsection{Resolving Input Dependencies}
Any intersection between the propagated set of consumed and produced elements of a loop indicates the possible presence of either a RAW (loop-carried) or WAR (input) dependency.
To identify an input dependency, we analyze the consumed and produced elements of a single loop iteration for a loop $\mathcal{L}$.
For any data container $D$ with a given symbolic offset expression $f$ inside the loop body's set of consumed elements,
we check if a write from a later loop iteration overlaps with it.
We do this by checking the offset expressions $g$ for each write to the same data container $D$ in the loop body's set of produced elements.
For any such produced offset $g$, an input dependency is present if and only if:
\[
\exists \delta > 0:\ f(\mathcal{L}_{var}) = g(\mathcal{L}_{var} + \delta \mathcal{L}_{stride})
\]
The presence or absence of such a $\delta$ can be identified by solving $f(\mathcal{L}_{var}) - g(\mathcal{L}_{var} + \delta \mathcal{L}_{stride}) = 0$ for $\delta$.
Note that due to the symbolic stride, this property also holds true for loops iterating in descending order or where the stride is a function of the loop variable itself.

If no other dependencies involve the data container $D$, the dependency can be resolved by creating a copy $D_{copy}$ of the container before the loop, and replacing reads from $D$ inside the loop body with reads from $D_{copy}$.
Only reads dominated by a write to the same offset in the loop body's dataflow graph can be left unchanged.
In our example from Fig.~\ref{fig:parallel-deps} this corresponds to replacing the read from \texttt{C} in \texttt{S2} with a read from \texttt{Cp} with the same offset, as demonstrated on the left in Fig.~\ref{fig:doacross-parallelization}.
Doing so removes the input dependency, since even in a parallel execution of the loop, each thread or iteration will read the same value as in the original, sequential scenario.

\subsection{Parallelization of Read-After-Write Dependencies}
After eliminating input and output dependencies, many loops can be parallelized with traditional DOALL parallelism.
For loops that still contain sequential dependencies, a pipeline (or DOACROSS) parallelism strategy can be explored.
In a fine-grained DOACROSS parallelized loop nest, each iteration of the loop may be run independently by a separate thread or process, but synchronization is used to ensure no read-after-write dependencies are violated.
For each statement with an access that depends on a specific other loop iteration, identified by some iteration vector $\vec{x}$, a \texttt{wait} statement that takes $\vec{x}$ as an argument can be inserted before the access in question.
This \texttt{wait} statement tells an executing thread or process to halt execution and yield control until a thread or process executing the loop iteration identified by $\vec{x}$ has reached a \texttt{release} statement.
The loop iterations are thus executed in a pipelined fashion, ensuring data dependencies are met before progressing in each iteration.
This flavor of pipeline parallelism can consequently be used in any parallel execution environment where the necessary communication and control granularity is available, such as in OpenMP or MPI.

\subsubsection{Determine Synchronization Points}
To determine at what points the loop iterations of a loop $\mathcal{L}$ need to be synchronized, the specific data accesses depending on previous loop iterations need to be identified together with the corresponding iteration vector.
Identifying these synchronization points functions analogously to how output dependencies are identified.
For any data container $D$ with a given symbolic offset expression $f$ in the loop body's set of consumed elements we check which write from a previous iteration overlaps with it.
We check for any offset expression $g$ belonging to a write to the same data container, where:
\[
\exists \delta > 0:\ f(\mathcal{L}_{var}) = g(\mathcal{L}_{var} - \delta \mathcal{L}_{stride})
\]

If solving $f(\mathcal{L}_{var}) - g(\mathcal{L}_{var} - \delta \mathcal{L}_{stride}) = 0$ for $\delta$, yields a result for $\delta$, the statement corresponding to that read depends on the iteration $\mathcal{L}_{var} - \delta \mathcal{L}_{stride}$ of that loop.
This check is performed for each of the loops in the loop nest attempting to be parallelized.
For any loop where no such $\delta$ exists, there is no dependency that can be synchronized with this strategy.
Similarly, if any data access exhibits one of the other types of dependencies and that dependency cannot be resolved, no parallelization is possible with this strategy.

Collecting the dependency offsets for all loops involved creates an iteration space vector of the form 
\[
(\mathcal{L}^{0}_{var} \pm \delta_0 \mathcal{L}^{0}_{stride}, \mathcal{L}^{1}_{var} \pm \delta_1 \mathcal{L}^{1}_{stride}, ..., \mathcal{L}^{m}_{var} \pm \delta_m \mathcal{L}^{m}_{stride})
\]
for each of the loops $0$ to $m$ involved in the dependency.
If no dependence on a particular iteration of loop $i$ is present, $\delta_i$ is 0.
The writes corresponding to the conflicting offset $g$ identified are recorded as statements that need to have been executed before any loop iterations depending on the write are allowed to continue.

In the example on the left-hand side of Fig.~\ref{fig:doacross-parallelization} we can identify the access to \texttt{B[i][k-1]} in statement \texttt{S1} as overlapping with the write to \texttt{B[i][k]} in statement \texttt{S3} with $\delta = 1$ in the \texttt{k}-loop.
No offset $\delta$ can be found for the \texttt{i}-loop, which means that statement \texttt{S1} carries a dependency with the iteration space vector $(k-1,i)$.
The $(k,i)$-th iteration will consequently need to block when reaching that statement, until the $(k-1,i)$-th iteration has passed statement \texttt{S3}.

\subsubsection{Synchronizing}
After identifying which statements need to be synchronized, there are potentially multiple dependencies across loop iterations.
To avoid excessive synchronization and to ensure a maximally large parallel part for the loop body, we want to ensure that any statements depending on previous iterations occur as late in the loop body as possible without breaking dependencies.
We can use the dataflow graph for the loop body constructed during the externally visible reads and writes analysis to perform such code motion while ensuring that no dependencies are violated.
Each access with dependencies is now prefixed with a \texttt{wait} statement, which waits for the \texttt{release} call of a specific iteration corresponding to the dependency's iteration vector offset.
The right-hand side of Fig.~\ref{fig:doacross-parallelization} shows how our example is augmented with a \texttt{wait} statement that resolves the RAW dependency, after performing code motion to reduce the size of the dependant region.

To insert dependency-resolving \texttt{release} calls, we check if any dependency resolving access in the dataflow graph of a loop iteration post-dominates all other resolving accesses.
If such a post-dominating access exists, a \texttt{release} is inserted after only that access to limit synchronization overhead.
If no resolving access post-dominates, waiting iterations are released at the end of the loop body's execution through a \texttt{release} statement at the end of the loop body.
If in addition to not having a post-dominating resolving access, a loop's first statement depends on a previous iteration, parallelization is skipped since there is no pipelining benefit to be extracted.

\section{Memory Schedules}\label{s:memory-schedules}
Even after reaching the maximal attainable degree of parallelism, a loop nest's performance is often still hindered by data movement costs.
Optimizations such as loop tiling and reordering, data layout modifications, and buffering on loop nests are thus impactful changes in an effort to improve caching behavior.
They reschedule computations to improve locality, and pin critical data containers to low latency cache.
While these optimizations aim to improve locality in programs and loop nests as a whole, there is still potential in optimizing on a fine-grained, per-access basis.

Leveraging the information provided by our symbolic loop stride and data access analysis, we provide a fine-grained \textbf{Memory Scheduling} pass.
The idea behind scheduling individual memory accesses, is to dictate when and how a specific access is performed, in an effort to have fine-grained control over the data movement associated with it.
A memory schedule can be thought of as a property of a data access that does not directly modify the IR of a program or the corresponding source code.
Instead it gets translated to specific patterns during lowering of the IR after optimization.
This is beneficial because fine-grained data movement optimizations such as the ones we discuss in this section fundamentally change the data accesses, which makes subsequent analyses for optimizations such as loop unrolling or vectorization more difficult.
By treating schedules merely as a property on an access instead, data access analysis and subsequent optimizations remain unaffected and the schedule is implemented during lowering.

\subsection{Automatic Software Prefetching}
Even when spatial and temporal data locality are exploited optimally, the number of cache misses is not reduced to zero due to initial cold misses.
This problem has long been understood, and modern processors avoid this by including hardware prefetching engines that attempt to predict what data will be used in the near future.
They achieve this by detecting patterns in the accesses made by an application and using those to predict future accesses, for which they asynchronously load data into the cache.
At the same time, compilers try to statically detect patterns such as strided array accesses so they can generate additional software prefetching instructions to improve the prefetch engine's accuracy.

Determining the correct data accesses to prefetch in the context of HPC applications can be difficult for both the prefetching engine and the compiler.
A variety of factors such as the use of irregular access patterns, short or large numbers of arrays, or not knowing the bounds of loop nests can make it difficult to predict what to prefetch~\cite{lee_when_2012, chen_performance_1994}.

\subsubsection{Access Prefetching Schedule}
To address this, we leverage SILO's loop representation to perform \textbf{Automatic Software Prefetching}.
Using knowledge about the application, i.e., about data accesses and surrounding loop characteristics, we determine which data accesses should be prefetched by marking them with a corresponding memory schedule.
This schedule translates to software prefetching instructions during lowering, helping the hardware prefetcher determine what to prefetch when and to what cache level, and whether to prepare the data for a read or write operation.

Incorrect or too frequent prefetching can be actively harmful, so it is best not to prefetch in inner-most loops.
Additionally, hardware prefetching units are able to prefetch data that is being accessed in a streaming manner, but are unable to anticipate sudden changes in a data access pattern~\cite{lee_when_2012}.

It is thus ideal to explicitly generate prefetching instructions any time such a sudden change in data access patterns occurs.
A data access pattern change is characterized by a sudden change in the strides between subsequent accesses.
As such, even if an array is padded for optimal alignment and uses parametric strides, hardware prefetching units may observe the access pattern and identify the correct strides after a few accesses, allowing them to prefetch correctly.
However, if the stride abruptly changes, such as in the case of tiled loops when transitioning between tiles, we can expect hardware prefetching units to incorrectly prefetch along the previous pattern until the new stride is identified.

\subsubsection{Prefetching Unpredictable Strides}
\begin{figure}
    \centering
    \includegraphics[width=.9\linewidth]{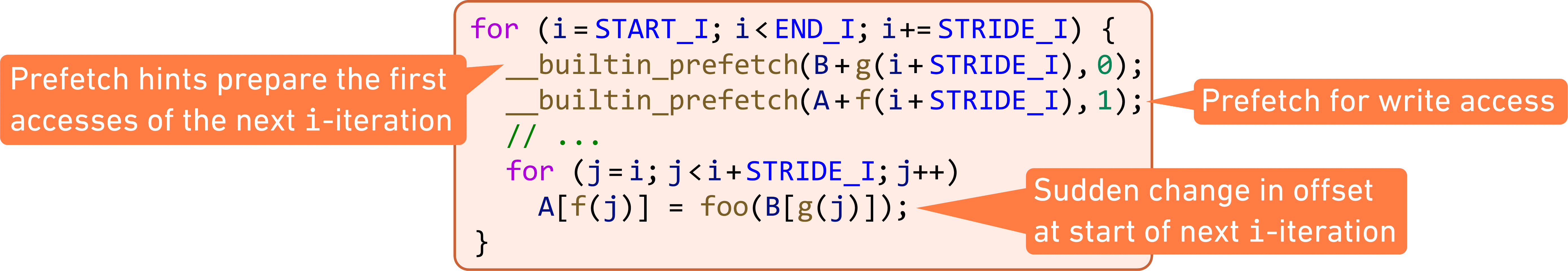}
    \vspace{-.3cm}
    \caption{Generating prefetch hints prepares prefetch unit for unexpected strides.}
    \Description{The code sample shown contains data accesses that depend on a loop stride, which in turn depends on an outer loop variable. This makes the next data access hard to predict. The code sample shows how inserting prefetch hints can mitigate this issue.}
    \label{fig:prefetch-hints}
\end{figure}

We can identify such sudden changes in strides in our loop representation any time a loop variable is being used in a data access, and that loop variable's starting value expression itself depends on the loop variable or a surrounding loop.
Such a scenario can be seen in Fig.~\ref{fig:prefetch-hints}, where the accesses to the arrays inside the loop depend on a loop variable that in turn depends on the surrounding loop variable \texttt{i}.
This pattern causes sudden stride changes in accesses to \texttt{A} between executions of the \texttt{j}-loop.

To help the prefetching unit with this jump, we can insert a prefetch instruction directly after the loop header of the surrounding \texttt{i}-loop to prepare the data for the next iteration of the \texttt{i}-loop.
The instruction must prefetch the involved data container (\texttt{A} in the example) offset to where the first access in the next loop iteration will occur.
This offset is obtained by identifying the first read to the data container inside the loop, and the corresponding access offset expression $f$.
Every occurrence of the loop variable \texttt{i} inside expression $f$ is then offset by the stride expression of the corresponding loop.
The resulting expression determines the offset of the prefetch instruction.
Depending on the type of data access (i.e., read or write), we indicate whether to prepare the data for a read or a write operation on the prefetch instruction.
If the loop or access in question depends on more than one other loop, the lowest one in the hierarchy (i.e., closest to the access) is chosen for the location of a prefetch instruction, since the immediate change in strides is brought on by the inner-most associated loop.
If a loop identifed for the insertion of a prefetch instruction is scheduled to be parallelized, a prefetch instruction is not beneficial and can be omitted.

These resulting prefetch instructions consequently never occur in the inner-most loop of a nest, causing them to be executed infrequently and only when an unpredictable change in access location occurs.
This means that unless the combined execution time of all loops further down the nest hierarchy is shorter than completion time of a single prefetch, the overhead of executing the prefetch operation is expected to be negligible~\cite{lee_when_2012}.

\subsection{Pointer Incrementation}
Even with good data locality and caching behavior, data values still must be loaded into registers to perform computations.
Since accessing registers is also the cheapest form of accessing data, it is critical for performance that values that may be reused are kept in register storage for as long as possible.
Unfortunately, the number of available registers in both CPUs and GPUs is low, with at most a few dozen available at a time.
This has the consequence that computations with many active intermediate values quickly run out of register space, causing values held in registers to be written back to other memory spaces (e.g., stack, cache, or off-chip memory) to make room for other values.
Such \emph{register spills} are expensive, as they cause data for subsequent computations to be read from memory again.
Register spills occurring inside of performance critical loop nests are consequently quickly a cause for performance degradation through an increase in data movement costs~\cite{chaitin_register_1982}.

Compilers have optimization passes that attempt to reduce the chances of register spills occurring and maximize register reuse.
However, even best practices documents of state of the art compilers~\cite{intel:opt-guide-gpu} are clear on the fact that compilers need the help of the developer to ensure that register spills are kept at a minimum.
Common optimization techniques employed to help the compiler analyze and optimize for this include:
\begin{itemize}
    \item Avoiding the use of large arrays or structures
    \item Reducing the number of live variables
    \item Selecting smaller sized data types
    \item Avoid loop unrolling (increases temporary variable count)
\end{itemize}

However, many of these techniques are not easily applicable in the context of  scientific computing applications or require intrusive global optimizations.
Particularly large, high-accuracy data types and large arrays are hard to avoid in real-world use cases.
Stencil computations, for instance, may access dozens of different indices of multiple large, multidimensional arrays (fields) in a single loop iteration, usually as a function of loop variables.
While the cost of performing the necessary offset arithmetic is typically negligible, the associated cost in register space is not.

Due to the complex, multi-dimensional access patterns, compilers often fail to reduce the register count sufficiently for these kinds of applications, causing spills that hinder performance.
These optimizations become even harder when pre- or post-padding has been introduced to arrays to improve alignment, in which case stride calculations increase the register count and the chances that reduction attempts fail.

To address this issue, we use the information provided by the symbolic loop stride and data access analysis to add a \textbf{Pointer Incrementation} memory schedule to specific data accesses.
Pointer incrementation is a form of loop strength reduction, which is the technique of replacing costly operations inside a loop with computationally less expensive ones.
In the case of accesses to multi-dimensional arrays, the costly operations are the array offset computations.
The less expensive alternative is to not offset the array pointer at each access, but instead increment it in accordance with the loop and array strides in each loop iteration.
Doing so allows the data accesses to be performed with constant offsets to a moving pointer.
This is not only significantly easier to analyze and optimize by the compiler, but can have such an impact on performance that there have even been attempts at developing specialized hardware for this optimization~\cite{schuiki_stream_2021}.

When setting a pointer incrementation schedule for a given data access we follow three basic steps:
Define a pointer that will be used to increment and access the data, increment it while iterating over the loop nest, and use it to replace the data access.
For each of these steps, respectively, we need to identify:
\begin{enumerate}
    \item where and to what value to initialize the pointer,
    \item where to increment the pointer and by what amount,
    \item the offset to the pointer used when accessing data.
\end{enumerate}

We go over each of these steps following an example shown in Fig.~\ref{fig:ptr-increment-process}, where a two dimensional array A $\in\mathbb{R}^{I \times J}$ is being accessed inside of a loop nest with two loops.

\subsubsection{Pointer Initialization}
For each data access where we want to perform pointer incrementation, we first identify each loop that impacts the offset computation for the given access.
We can analyze the symbolic expression that determines the access offset and determine any loop $\mathcal{L}$ surrounding the access to be involved, if the offset expression relies on its iteration variable $\mathcal{L}_{var}$.
If that is the case, that loop $\mathcal{L}$ has a direct effect on the computed offset and needs to be considered when incrementing a pointer.

\begin{figure}
    \centering
    \includegraphics[width=.95\linewidth]{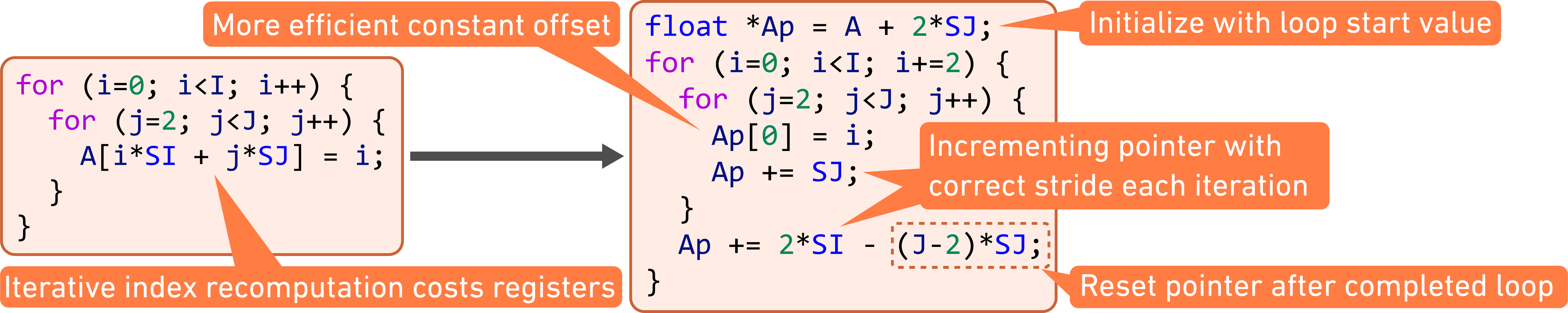}
    \vspace{-.3cm}
    \caption{Using pointer incrementation saves expensive index recomputations.}
    \Description{The image shows a code sample with two nested loops and a strided data access inside. The code sample is transformed to not perform a strided data access with index recomputation in each loop iteration, but instead carry a pointer that is used to access the data with a single offset, and increment the pointer in each loop iteration accordingly.}
    \label{fig:ptr-increment-process}
\end{figure}

The pointer definition and initialization must live outside of the loop nest it is being used in, so the outermost loop involved in the data access is identified, and the definition is emitted directly before that loop header.
To avoid data races with incrementing pointers across parallel loop executions, pointer definitions are emitted above the top-most sequentially scheduled loop when inside of a loop nest with parallel loops.

The pointer is initialized to the base pointer to the data container $D$ being accessed, offset by the symbolic indexing expression $f$.
For any involved loops \emph{below} the pointer initialization in the loop nest hierarchy, the symbol corresponding to their loop variable in the indexing expression is replaced with the symbolic expression that sets the starting point of that loop.
In the example shown in Fig.~\ref{fig:ptr-increment-process} this corresponds to replacing the symbols \texttt{i} in the indexing expression with $0$ for the start of the \texttt{i}-loop, and the symbol \texttt{j} with $2$ for the start of the \texttt{j}-loop.

\subsubsection{Pointer Incrementation}
To ensure the pointer remains aligned to the correct index in the array, it now needs to be incremented with the completion of each loop iteration.
This ensures that given an iteration vector $\vec{x}$ characterizing a specific point in the iteration space by giving each iteration variable a value, the pointer points to the same location as given by $D[f(\vec{x})]$.
To achieve this, we need to determine the amount by which the pointer needs to be incremented per iteration for each loop in a loop nest.
Given a loop $\mathcal{L}$ for which the symbol $\mathcal{L}_{var}$ can be found in the offset expression $f$ of $D$, the corresponding pointer $D^p$ must be incremented by $\Delta^{D^p}_i$, which is the difference in the offset to the container between the current and subsequent loop iterations.
$\Delta^{D^p}_i$ can thus be calculated as
\[
\Delta^{D^p}_{i} = f(\mathcal{L}_{var} + \mathcal{L}_{stride}) - f(\mathcal{L}_{var})
\]

In case two or more loops are involved, the pointer must also be reset again each time any involved loop finishes executing, except for the outer-most loop.
This equates to subtracting every incrementation $\Delta^{D^p}_i$ from the pointer that was made during execution of the loop.
For loop $\mathcal{L}$, the amount $\Delta^{D^p}_r$ to subtract from the pointer after execution thus equates to multiplying the number of iterations by $\Delta^{D^p}_i$ for that loop, or more generally to
\[
\Delta^{D^p}_{r} = f(\mathcal{L}_{end}) - f(\mathcal{L}_{start})
\]
Using symbolic arithmetic, both resulting symbolic expressions $\Delta^{D^p}_{i}$ and $\Delta^{D^p}_{i}$ can now be simplified statically.
Any time $\Delta^{D^p}_{r}$ for a given loop is symbolically equal to $\Delta^{D^p}_{i}$ of a surrounding parent loop, both the reset and subsequent incrementation in the outer surrounding loop can be omitted, saving further register space.

In the example from Fig.~\ref{fig:ptr-increment-process}, after simplifying both expressions, this amounts to incrementing the pointer by \texttt{SJ} for each iteration of the \texttt{j}-loop, and by $2 * \texttt{SI}$ for each iteration of the \texttt{i}-loop.
In addition to that, after each completion of the \texttt{j}-loop the pointer needs to be reset by $(\texttt{J} - 2) * \texttt{SJ}$ to reverse all incrementations performed by the \texttt{j}-loop.

\subsubsection{Determining Pointer Offset}
By incrementing pointers in this way, the offset used when accessing the array through them equates to 0 in most cases, since the pointer was initialized with any constant offsets already included.
However, in some cases the same array is accessed multiple times inside the same loop with a distance that can be expressed as a compile-time constant $\delta$.
In such a case, we can free up further registers by using only one pointer to increment for all accesses, and performing accesses with the constant offset $\delta$ on the pointer.
This allows the compiler to perform multiple accesses with no added register cost.

\section{Implementing SILO}\label{s:implementation}
To implement SILO according to the architecture outlined in Sec.~\ref{ss:architecture}, two main components are required from a targeted HPC framework or compiler:
\begin{enumerate}
    \item An IR capable of symbolic and dataflow analysis
    \item A way of providing custom code lowering rules to implement memory schedules and DOACROSS parallelism
\end{enumerate}
Both of these components are available in many modern HPC frameworks and compilers, such as LLVM~\cite{lattner_llvm_2004} through MLIR dialects~\cite{lattner_mlir_2021}, polyhedral representations~\cite{grosser_polly-polyhedral_2011, bondhugula_model_2010}, or dataflow programming frameworks such as HPVM~\cite{kotsifakou_hpvm_2018} or DaCe~\cite{ben-nun_stateful_2019}.

We provide a prototype implementation to interact with the DaCe framework because the nature of its dataflow IR simplifies the necessary data dependency analysis for data privatization.
The same data dependency resolution also requires solving symbolic equations.
We use the symbolic mathematics library SymPy~\cite{sympmy} to handle symbolic expressions and solve equations.
To evaluate DOACROSS parallelization, we provide custom lowering rules to generate code for shared-memory parallelism using OpenMP, with synchronization primitives introduced in the OpenMP 4.5 specifications~\cite{rendell_expressing_2013, omp:4.5-specification}.

\begin{figure}
    \centering
    \includegraphics[width=.95\linewidth]{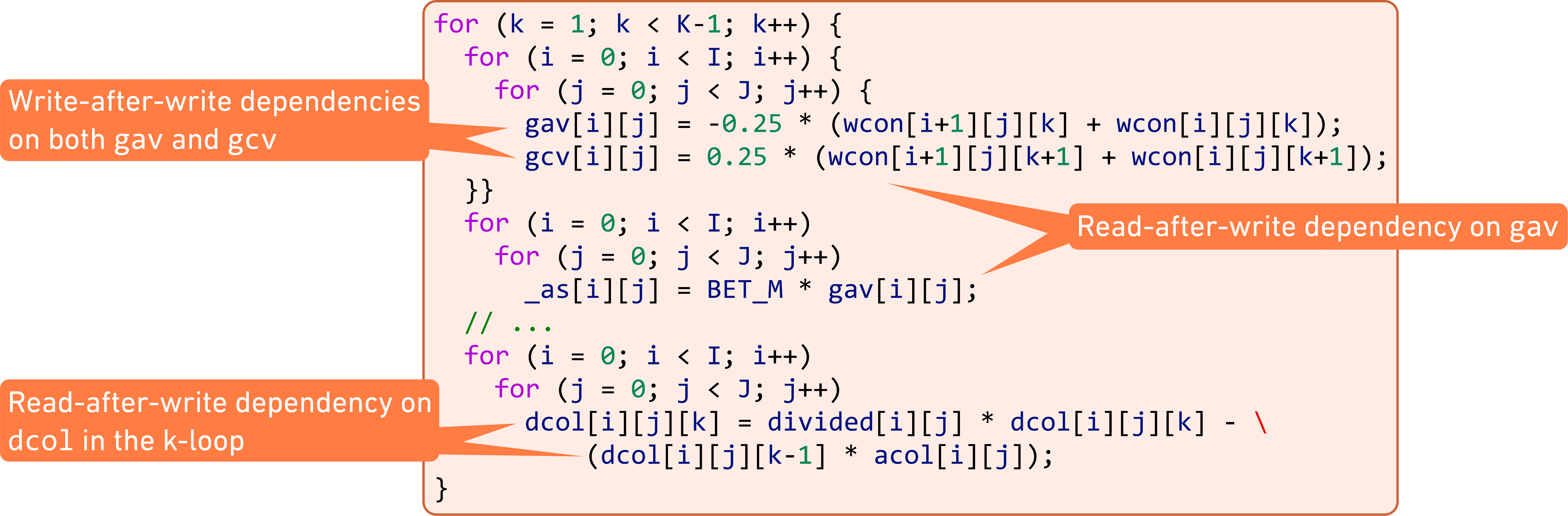}
    \vspace{-.3cm}
    \caption{Loop-carried dependencies preventing parallelization over the \texttt{k}-loop.}
    \Description{The image shows a code sample depicting various loop-carried dependencies that prevent parallelization of the k-loop in vertical advection.}
    \label{fig:vadv-code}
\end{figure}

\section{Evaluation}
We use our prototype implementation to automatically optimize a real-world atmoshperic modeling application and a series of fundamental scientific compute kernels, and compare against state of the art compilers and optimization frameworks.
For all our experiments we use gcc 10.2.0, clang 15.0.7, and icc 2021.3.0.
We run DaCe 0.15.1 with Python 3.12.2, Pluto 0.12.0-2-g001cb37, and Polly with LLVM 15.0.7, which additionally performs scalar evolution.
All of our experiments are performed on two nodes.
One node is equipped with two 18-core Intel Xeon Gold 6140 at 2.3 GHz with 768 GiB of RAM, and the other node is equipped with two 64-core AMD EPYC 7742 at 2.25 GHz with 512 GiB of RAM.

\subsection{Parallelizing Sequentially Dependent Loops}
Vertical Advection is a tridiagonal solver used in weather and climate models, which solves advection equations in a 3-dimensional ($I \times J \times K$) domain with the Thomas algorithm.
While the two horizontal dimensions ($I \times J$) can be trivially parallelized with DOALL parallelism, this leads to a series of sequential stages in the third, vertical dimension ($K$).
We use an implementation of the solver from the High-Performance NumPy Benchmarking suite NPBench~\cite{ziogas_npbench_2021} and use it to extract parallelism from the vertical dimension.

The implementation contains two runtime-dominating loop nests: One to perform a forward sweep of the Thomas algorithm, and one to perform backwards substitution.
Both loop nests consist of a sequential outer loop in the $K$ dimension with a series of two-loop nests over $I$ and $J$ inside, each of which may be fully parallelized with DOALL parallelism.
However, parallelizing the $K$ dimension is not trivially possible due to a series of WAW and RAW dependencies between loop iterations.
An excerpt of the forward sweep loop nest is shown in Fig.~\ref{fig:vadv-code}, where both types of dependencies are visible.

\begin{figure*}
    \centering
    \begin{subfigure}{.46\linewidth}
        \centering
        \includegraphics[width=.75\linewidth]{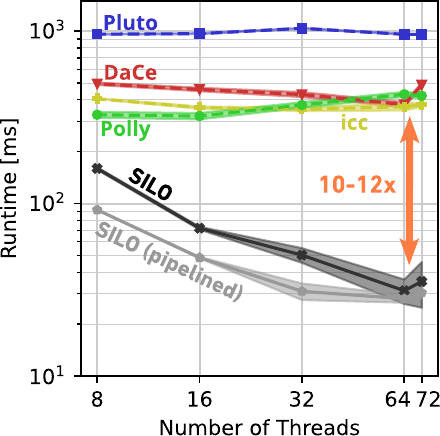}
        \caption{Strong scaling on the node with two Intel Xeon Gold 6140.}
        \label{fig:vadv-ss-ault6}
    \end{subfigure}
    \begin{subfigure}{.46\linewidth}
        \centering
        \includegraphics[width=.75\linewidth]{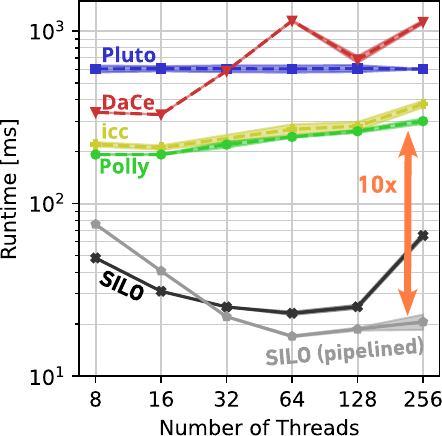}
        \caption{Strong scaling on the node with two AMD EPYC 7742.}
        \label{fig:vadv-ss-ault19}
    \end{subfigure}\\\vspace{.15cm}
    \begin{subfigure}{.46\linewidth}
        \centering
        \includegraphics[width=.94\linewidth]{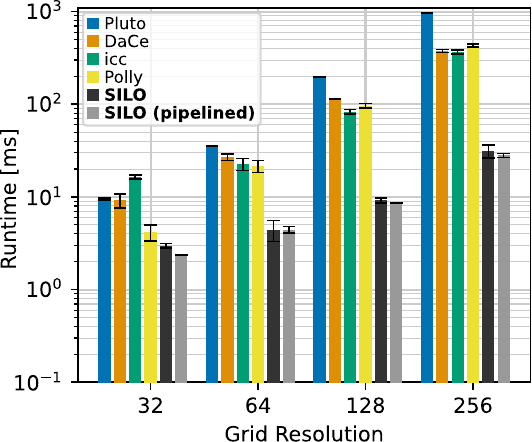}
        \caption{Varying grid size with 64 threads (Intel node).}
        \label{fig:vadv-inc-ault6}
    \end{subfigure}
    \begin{subfigure}{.46\linewidth}
        \centering
        \includegraphics[width=.9\linewidth]{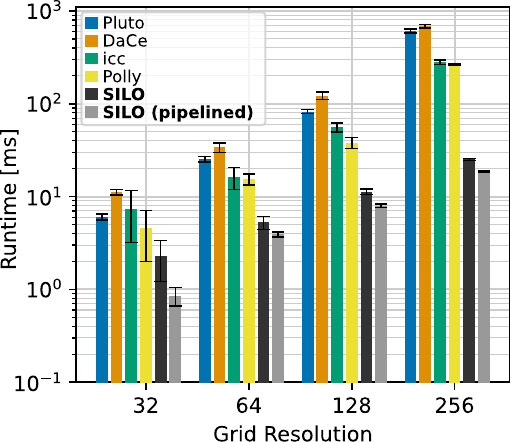}
        \caption{Varying grid size with 128 thread (AMD node).}
        \label{fig:vadv-inc-ault19}
    \end{subfigure}
    \vspace{-.3cm}
    \caption{SILO automatically pipelines sequentially dependent loop nests in vertical advection, improving performance and strong scaling over state of the art optimizers. All plots use $K=180$ in the vertical dimension. Strong scaling is shown for a $256 \times 256$ grid.}
    \Description{The figure shows four images: the first showing strong scaling on an intel node, the second showing strong scaling on an AMD node, the third showing runtime with respect to varying grid resolutions on an intel node, and the fourth showing the same on an AMD node.}
    \label{fig:vadv-results}
    \vspace{-.3cm}
\end{figure*}

We use Polly, Pluto, and DaCe, as well as our implementation of SILO to automatically extract as much parallelism from the application as possible.
To apply Polly and Pluto, we provide a C implementation of the kernel, ensuring that pointers are annotated with \texttt{restrict} to indicate no aliasing, annotating the kernel as a static control part (SCoP), and using a compatible multidimensional array notation shown in Fig.~\ref{fig:vadv-code}.
Polly was run with parallelization enabled, and Pluto was run with the flags \texttt{--parallel --multipar}, which gave the best results.
For DaCe, we use the auto optimization pass integrated in the NPBench implementation.
The same implementation is used for SILO, where we run two optimization configurations:
Configuration 1 is used to eliminate sequential dependencies where possible, before handing back over to DaCe's auto optimization.
Configuration 2 builds on configuration 1, but additionally applies automatic pipelining to loops with read-after-write dependencies.
An overview of the measured runtimes can be seen in Fig.~\ref{fig:vadv-results}.

Polly and Pluto both detect the entire kernel as one contiguous SCoP and perform various amounts of unrolling, vectorization, and tiling on the loop nests.
However, due to the presence of WAW conflicts when parallelizing across the $K$ dimension, they are both unable to parallelize all available dimensions.
Intel's icc similarly parallelizes a part of the application, but reports no parallelization across the $K$ dimension due to dependencies.
DaCe fails to perform any tiling or vectorization, but fuses many loops together, which results in some arrays being converted to temporary scalars.
While removing some arrays, this does not eliminate all WAW and RAW dependencies, and consequently DaCe only extracts parallelism across the $I$ and $J$ dimensions, which still reside inside the sequential $K$ loop.

Running SILO with configuration 1 eliminates all WAW dependencies present in the loop nests, which allows the automatic optimization to move the $K$ loops inside of the $I$ and $J$ loops in a subsequent pass.
While this still leaves the loop sequential, it already leads to a significant \textbf{speedup of up to 10$\times$} over Polly, the fastest previous result.
With configuration 2 SILO additionally parallelizes the sequential $K$ loop in the forward sweep phase of the application by inserting the necessary synchronizations.
Fig.~\ref{fig:vadv-inc-ault6} and~\ref{fig:vadv-inc-ault19} show how parallelizing across all three dimensions increases the speedup over Polly to \textbf{up to 12$\times$}.
The additional parallel dimension brings the most speedup for small problem sizes or for large numbers of threads, where the speedup over configuration 1 can go up to a factor of 2.7$\times$.
Fig.~\ref{fig:vadv-ss-ault6} and~\ref{fig:vadv-ss-ault19} also indicate how this additional degree of parallelism results in improved strong scaling.

\subsection{Automatic Software Prefetching}
To demonstrate the effects of automatic software prefetching, we use SILO in conjunction with DaCe to optimize a matrix-matrix multiplication for two double precision square matrices of size $4096 \times 4096$.
Using an optimization recipe provided by DaCe, we first let it optimize a Python implementation of the matrix-matrix multiplication, which serves as our baseline.
The recipe tiles the matrix multiplication twice, creating a buffer for the tile of the output matrix, and a buffer for one of the input matrices.

\begin{table}[]
    \centering
    \caption{\textup{Effect of prefetching on a tiled matrix multiplication.}}
    \vspace{-.3cm}
    \addtolength{\tabcolsep}{-0.2em}
    \begin{tabular}{lrrrr}
        \toprule
        \small{\textbf{Compiler}} & \multicolumn{4}{c}{\small{\textbf{System}}} \\
        \cmidrule(lr){2-5}
        & \multicolumn{2}{c}{\small{Intel Node}} & \multicolumn{2}{c}{\small{AMD Node}} \\
        \cmidrule(lr){2-3}
        \cmidrule(lr){4-5}
        & \small{No Prefetch} & \small{Prefetching} & \small{No Prefetch} & \small{Prefetching} \\
        \midrule
        \small{gcc} & \small{752.34 ms} & \small{549.53 ms} & \small{151.22 ms} & \small{149.32 ms} \\
        \small{clang} & \small{273.96 ms} & \small{266.65 ms} & \small{99.05 ms} & \small{90.98 ms} \\
        \small{icc} & \small{412.41 ms} & \small{411.80 ms} & \small{552.94 ms} & \small{553.58 ms} \\
        \midrule
        & \small{Intel MKL} & \small{116.80 ms} & \small{OpenBLAS} & \small{122.47 ms} \\
        \bottomrule
    \end{tabular}
    \label{tab:prefetch-res}
\end{table}

To assess the effects of prefetching, we take this optimized matrix multiplication and apply SILO's memory scheduling pass.
We compare the runtime before and after generating prefetch hints with SILO when compiling with three different compilers, namely gcc, clang, and Intel's icc.
An overview of the results is shown in Table~\ref{tab:prefetch-res}.
The biggest improvement through prefetching can be seen for gcc on the Intel system, where application informed prefetching leads to a speedup of \textbf{1.37$\times$}.
Interestingly, there is nearly no observed speedup for gcc on the AMD system, with less than 1\% difference in runtime.
With clang we observe a speedup of 1.03$\times$ and 1.09$\times$ on the Intel and AMD systems, respectively, and icc seems to not benefit noticeably from additional prefetch hints, while performing noticeably poor overall on the AMD system.

\begin{figure*}
    \centering
    \includegraphics[width=.75\linewidth]{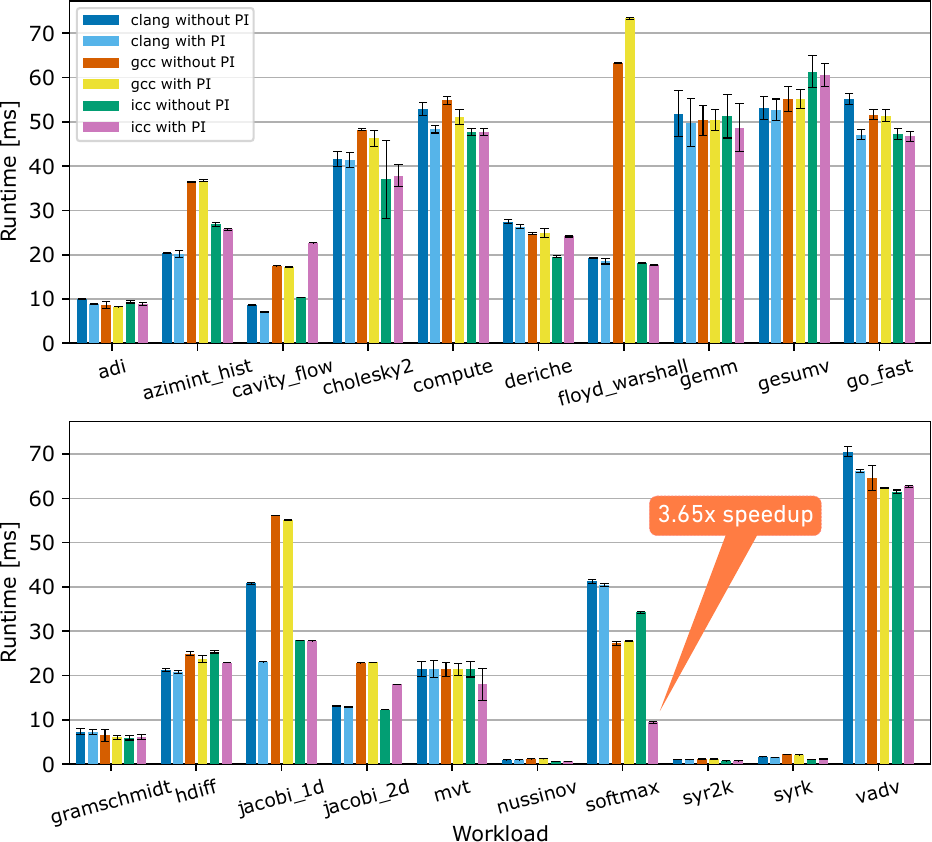}
    \vspace{-.3cm}
    \caption{Effects of pointer incrementation on the NPBench benchmark suite.}
    \Description{The figure shows the effect of pointer incrementation on the NPBench benchmark suite, showing how PI consistently either has no effect or significantly improves performance.}
    \label{fig:npbench-res}
\end{figure*}

\subsection{Pointer Incrementation}
We optimize 34 benchmarks\footnote{Remaining 16 omitted due to DaCe 0.15.1 crashing when handling them.} from the NPBench benchmarking suite~\cite{ziogas_npbench_2021} using DaCe's automatic optimization without our added parallelization pass.
Using SILO we then schedule all memory accesses to arrays inside of loops with pointer incrementation.
For compiling the resulting optimized code we used clang, gcc, and icc.
Each benchmark is run with the medium data preset from NPBench, and in each case we record the times before and after scheduling memory accesses with pointer incrementation and report the speedup.
We give an overview of the resulting runtimes for 20 of these benchmarks in Fig.~\ref{fig:npbench-res}.
The remaining 14 are not listed because DaCe replaces the majority of the benchmark with library calls, rendering them unaffected by our further optimizations.

Out of all the shown benchmark and compiler combinations ($20 \times 3$), 30 showed a noticeable ($>\pm3\%$) change in performance.
27 of the benchmark and compiler combinations improved in performance with an average of 1.20$\times$.
Notable examples are the jacobi\_1d benchmark, which improved by a factor of 1.76$\times$ when compiled with clang, and softmax, where we observe \textbf{a 3.62$\times$ speedup} when compiled with icc.

\section{Related Work}
Due to their central role in scientific computing and applications in general, loops and optimizations around them have been the center of decades of research.
We highlight some of the works most closely related to SILO and how they differ from our approach in the following.

\subsection{Symbolic Loop Analysis}
Symbolic analysis has been used to a certain extent in optimizing compilers to analyze data dependencies and perform program simplifications~\cite{blume_overview_1994, goos_demand-driven_1996, scholz_unified_2000}.
It is frequently used for generalized strength reduction inside of loops, to enable loop invariant code motion, and to perform dependence tests~\cite{goos_symbolic_1993, pottenger_idiom_1995, van_engelen_unified_2004}.
Similarly, tools such as Polly~\cite{grosser_polly-polyhedral_2011}, Pluto~\cite{bondhugula_pluto_2016}, or CHiLL~\cite{chen_chill_2008} use the polyhedral model to analyze loops to extract parallelism and improve data locality, in an approach similar to the symbolic analyses discussed here.
There are also techniques to alleviate the constraint of affine expressions with the polyhedral model~\cite{goos_symbolic_2004, hutchison_polyhedral_2010}.

However, current tools built on the polyhedral model focus on schedule changes for loops to improve parallelism and data locality, which potentially leaves hidden parallelism unexploited.
By also considering data allocation changes, SILO fills this gap to expose and exploit more degrees of parallelism where available.

Aside from Polly, LLVM~\cite{lattner_llvm_2004} also makes use of symbolic analysis through an analysis pass called scalar evolution, which analyses chains of recurrences~\cite{bachmann_chains_1994, van_engelen_unified_2004} to deduce how scalar values change over time.
This information is used by LLVM to perform crucial optimizations such as loop strength reduction.

SILO extends the class of analyzeable loops and loop body statements considerably by applying to multi-level loop nests and multidimensional arrays.

Symbolic analysis has also been employed in domain- or purpose-specific tools and techniques.
For example, Chikin et al.~\cite{chikin_memory-access-aware_2019} use it to evaluate profitable loop transformations for GPU-bound OpenMP applications.
Sympiler~\cite{cheshmi_sympiler_2017}, GLU~\cite{he_gpu-accelerated_2016}, and ParSy~\cite{cheshmi_parsy_2018} use symbolic analysis to parallelize and optimize sparse matrix computations.
SILO uses similar analysis techniques for generalizing optimizations of loop nest applications in scientific computing.

\subsection{Parallelization and Dependencies}
Maximizing parallelism in scientific computing applications is also one of the goals of polyhedral tools such as Polly~\cite{grosser_polly-polyhedral_2011}, Pluto~\cite{bondhugula_pluto_2016}, and PPCG~\cite{verdoolaege_polyhedral_2013}.
There, resolving read-after-write dependencies is typically achieved through wavefront parallelism where possible with techniques like diamond tiling~\cite{bondhugula_diamond_2017, liu_revisiting_2019}.
Alternative pipeline pattern detection techniques have been explored by Talaashrafi et al.~\cite{talaashrafi_pipeline_2022} to exploit parallelism between two or more loops.
A similar technique was recently presented by Cheshmi et al.~\cite{cheshmi_runtime_2023} for fusing two sparse matrix kernels with sequential dependencies.
On the other hand, parallelizing such dependencies with synchronizations has been explored in Fortran compilers by Midkiff and Padua~\cite{midkiff_compiler_1987} and by Unnikrishnan et al.~\cite{hutchison_practical_2012} through a technique called dependence folding~\cite{hutchison_practical_2012}.
These techniques are often complicated by strided accesses such as found in atmospheric modeling applications, which is why domain specific approaches have been presented in the past, such as by Maleki and Burtscher~\cite{maleki_automatic_2018}.

The symbolic loop and data access analysis employed by SILO addresses this by generalizing the required fine-grained data dependence analysis, including to expressions not represented in the polyhedral framework.

\subsection{Fine-Grained Data Movement Optimizations}
Prefetching with scientific computing applications is known to be difficult~\cite{jarvis_characterizing_2014, lee_when_2012, marin_diagnosis_2013}, which is why various solutions have been put forth to perform software prefetching for specific applications.
Hadade et al.~\cite{hadade_software_2020} have developed an approach to perform software prefetching for unstructured mesh applications.
Ainsworth and Jones~\cite{ainsworth_software_2017} similarly present a technique for prefetching in applications with indirect memory accesses.
For stencil applications, a model-based prefetching approach was put forth by de la Cruz and Araya-Polo~\cite{jarvis_modeling_2015}.

Similarly, register spills are a well known problem, which is why a lot of work has gone into finding ideal register allocation strategies for compilers~\cite{bernstein_spill_1989, de_moor_register_2009, appel_optimal_2001, chaitin_register_1982}, often with the help of graph coloring.
Motwani et al.~\cite{motwani_combining_1995} have shown that finding an optimal combined register allocation and instruction scheduling solution is NP-hard, and propose an approximate solution.
Because the problem is particularly challenging for scientific computing applications, multiple solutions have been proposed for specific situations, or specific architectures~\cite{salgado_register_2014}.
Examples include techniques specific to stencil programs~\cite{rawat_register_2018} and GPU applications~\cite{sakdhnagool_regdem_2019}.

With SILO we lift these optimizations from the compiler to a higher level representation where the enabled symbolic analysis can address the issues for general HPC applications with multi-level loop nests.

\section{Conclusion}
We develop symbolic, inductive loop analysis, a symbolic loop and data access analysis framework to improve automatic loop optimization capabilities in HPC applications.
Empowered by this analysis framework, we establish automatic optimizations to maximize parallelism and reduce data movement associated with loops.
We employ data privatization and automated fine-grained pipelining with synchronization to exploit more degrees of parallelism, and introduce \emph{memory schedules} to improve prefetching and register allocation.
With a prototype implementation of SILO we observe consistent speedup on various HPC benchmark applications through improved register allocation, and up to 12$\times$ speedup for a real-world atmospheric modeling application through pipeline parallelism.

Aside from our prototype implementation, the techniques outlined in this paper are not bound to a specific framework or program representation.
An implementation of SILO in a polyhedral framework for example is possible, though it would be limited to analyzing affine loop strides and unable to analyze data dependencies outside of a loop nest, restricting privatization.
The capabilities offered by the inductive loop analysis framework introduced with SILO can further be used for various other difficult-to-generalize loop optimization techniques.
With possible optimizations including the detection of computations and scans that can be represented with collective operations such as \texttt{MPI\_Scan}, the analyses presented in this paper can help optimizing frameworks get more out of loop nests in HPC applications.

\begin{acks}
This project has received funding from the European Research Council (ERC) under the European Union’s Horizon 2020 program (grant agreement PSAP, No. 101002047), and from the European High Performance Computing Joint Undertaking (EuroHPC-JU) under the DEEP-SEA program (grant agreement No. 955606).
Work by Lawrence Livermore National Laboratory was performed under the auspices of the U.S. Department of Energy under contract DE-AC52-07NA27344 (LLNL-JRNL-2013052).
\end{acks}

\bibliographystyle{ACM-Reference-Format}
\bibliography{zotero, additional}

\end{document}